# Connectivity Analysis of Directed Highway VANETs using Graph Theory

Samar Elaraby, and Sherif M. Abuelenin*

*Corresponding author. Department of Electrical Engineering, Faculty of Engineering, Port-Said University, Port-Fouad, Port-Said, 42526, Egypt. E-mail: s.abuelenin@eng.psu.edu.eg

**Abstract**—Graph theory is a promising approach in handling the problem of estimating the connectivity probability of vehicular ad-hoc networks (VANETs). With a communication network represented as graph, graph connectivity indicators become valid for connectivity analysis of communication networks as well. In this article, we discuss two different graph-based methods for VANETs connectivity analysis showing that they capture the same behavior as estimated using probabilistic models. The study is, then, extended to include the case of directed VANETs, resulting from the utilization of different communication ranges by different vehicles. Overall, the graph-based methods prove a robust performance, as they can be simply diversified into scenarios that are too complex to acquire a rigid probabilistic model for them.

## I. Introduction

Vehicular communications play an essential role in intelligent transportation systems[1], with applications varying from safety and driver assistant systems to providing infotainment and internet access[2,3]. In vehicular communications, vehicles are equipped with dedicated short-range communication (DSRC) devices to enable information exchange among them[4]. Vehicular ad-hoc networks (VANETs) are a special type of mobile ad-hoc networks (MANETs) in which, information is exchanged spontaneously among vehicles. Unlike other MANETs, the nodes in VANETs (i.e. vehicles) are highly mobile[4] resulting in a rather fast changing network topology. The movement of the nodes is confined to roads, which restricts their directions of motion, and the distances between them follow certain spatial distributions that are well studied and modeled within the framework of traffic theory. These characteristics result in high probability of network partitioning and no guarantee of end-to-end connectivity[5].

Studying connectivity features is a key aspect in analyzing and planning vehicular networks[6–8]. In general, a wireless ad-hoc network is called connected if a communication path (through single- or multiple hops) exists between any two nodes in the network[9,10]. In other words, the network is connected if, at the considered time instant, all of its nodes form a single cluster with no isolated nodes. Other definitions for connectivity were also introduced. In routing[11], delivering the packets towards a destination node has priority over reaching all vehicles. Thus, the aim is to find the optimal route between source and destination vehicles using strategies like the most strongly connected route strategy[8]. The connectivity is defined in such case to be the existence of a multi-hop route between the source and destination vehicles. In delay-tolerant applications[12–15], another strategy can be followed based on carry-and-forward communications[6]. In such networks, the packets are forwarded to vehicles that store and carry the data until they meet the recipient vehicle.

VANETs connectivity analyses have mostly relied on the assertion that the communication link between any two vehicles in the network is bidirectional[6,16–24], i.e. if one vehicle can communicate with another one then the opposite is true. This is based on the assumption that vehicles in a VANET would utilize similar communication ranges and on the reciprocity of the communication channel. Several approaches were introduced in estimating the connectivity probability in such cases including graph theory based simulations[6,25,26]. To the best of our knowledge, very few articles have considered the problem of estimating the connectivity of directed VANETs. A directed VANET topology can arise, in principle, for different reasons, for example, when two vehicles exchange data packets using two different carrier frequencies. But, more practically, when different vehicles utilize different communication ranges (i.e. different transmit powers). Analytic estimation of the connectivity probability of directed networks is not an easy tasks and requires introducing approximations[27,28] which could affect the estimation accuracy.

In this paper, we use tools from graph theory in simulating and estimating the instantaneous connectivity probability of directed highway VANETs resulting when different communication ranges are employed by different vehicles. Graph theory is the mathematical framework of complex network science. A network (any networked system) can be considered as a diagrammatical representation of some physical structure with interconnections between its elements (nodes)[29]. A graph, on the other hand, is a mathematical notion that represents only the structure of the considered network[29]. Graph theory-based modelling of wireless networks captures the network topology in matrix format and hence enables accurate and robust numerical simulations of network behavior. Network performance metrics including connectivity can be deduced in a straightforward manner from the graph-based models. Random geometric graphs[30] can easily capture the distance dependence and randomness in the connectivity of the nodes in comparison to other analytical methods. Two different graph-based simulation algorithms are discussed, namely, the Laplacian-eigenvalues and the adjacency-exponent approaches. The two methods were previously applied in studying undirected networks[25,26]. We review the two methods and compare the connectivity estimation results of both approaches to the analytic results while all vehicles utilize a fixed communication range. We discuss the suitability of each of the two methods in analyzing the different designations of connectivity. We show that the major difference between both methods is that the Laplacian-based method determines whether all vehicles are connected altogether as a single mesh while the adjacency-exponent approach determines whether there exists a path between the two end-vehicles. Afterwards, the algorithms are extended to more sophisticated scenarios, when the employed communication ranges vary among different vehicles, resulting in directed vehicular networks. We begin by considering the case of randomly employing two different communication ranges, then we extend our analysis to the case where the communication ranges of the vehicles follow uniform random distribution. While the adjacency-exponent method can be directly applied to analyzing connectivity of directed networks, we propose a tweak to the adjacency-matrix method to enable its use in such situations. The graph theory-based algorithms are shown efficient in capturing the network connectivity in such scenarios where probabilistic modeling is challenging. The paper is organized as follows. In Section II, we review related work including the notion of instantaneous connectivity and the basics of the representation of a wireless network as a graph. Section III discusses the connectivity of highway VANETs under the fixed communication range assumption. In section IV, we apply the graph theory tools to analyze the considered VANET connectivity. Then, we propose a modification to the graph representation of VANETs in order to extend the method to the case of different communication ranges in Section V. The paper is concluded in section VI.

**II. Related Work**

*A. Connectivity Probability*

Connectivity is a fundamental requirement in planning VANETs[23]. A wireless ad-hoc network is called connected if a communication path (single- or multi-hop route) exists between any two nodes in the network[9,10]. In other words, the network is connected if, at the considered time instant, all of its nodes form a single cluster with no isolated nodes. A measure of connectivity in VANETs is the connectivity probability, i.e. the probability that the network forms a single connected component[10]. The two terms, *connectivity* and *connectivity probability*, have been used interchangeably in literature[31,32]. Knowing the traffic headway distribution for given traffic conditions allows estimating the connectivity probability as a function of the node density and the communication range. This probability may be determined directly[16,17,24,33] or by finding the probability of network clustering first (e.g.[9]). Estimating the connectivity probability as a function of these parameters permits finding the minimum communication range (or alternatively, the minimum transmit power[34]) required for almost guaranteed connectivity.

Due to mobility, network topology changes with time, which in turn changes the connectivity continuously. Researchers[6,16,22,24,33,35] have relied on the instantaneous connectivity analysis, which estimates the probability that a given network is connected at a given time instance, e.g. the times of data-packet transfer[6]. The time dependence is dropped because instantaneous connectivity is used to study the average behavior of the network under certain conditions. The average network behavior is studied through averaging the network connectivity of each considered realization (e.g. 1 for connected and 0 for disconnected network) over large number of network realizations under the same conditions. These conditions include node density, communication range, spatial distribution of nodes, etc. This is complemented by studying the connection duration, as a function of the network mobility[8,22,36,37] parameters, such as vehicles velocity and the changing headway distributions.

A very common method of estimating the connectivity is to use the unit disc model[3,6,16,17,22,35,38,39] (i.e. fixed transmission range model). The model assumes that the coverage of each vehicle is a circle, with the vehicle at its center and the radius of the circle equals the vehicle's transmission range[38], $R$. It has been commonly assumed that of all vehicles within the network utilize the same transmission range. If the distance between any two vehicles is less than $R$, they are connected.

## B. Graph Theory Analysis of Networks

Graph theory is the mathematical framework used in studying complex networks. Any networked system (i.e. a set of interconnected nodes) can be represented as a graph $G(\mathbf{E},\mathbf{V})$ in which, the set of vertices **V** represent nodes and the set of edges **E** represent the links between nodes. Different tools from graph theory (or more generally, complex network science) have been used in modelling and studying different wireless networks[6,25,26,32,39–49]. In dynamic networks such as MANETs, both **E** and **V** are functions of time, and following the notation of the instantaneous connectivity, the average behavior is studied by considering the instantaneous graphs that represents topologies of sampled instants of the network. The different graph realizations can be obtained either by randomly generating them following certain constrains to model realistic scenarios, or through obtaining real traffic streams data[6,39,50]. To generate the random realizations of VANET graphs, each edge $e_{ij} \in \mathbf{E}$ (the edge between the pair of vertices $(v_i, v_j)$) is assigned with a probability that is a function of the physical distance and the communication channel characteristics between the two nodes[32].

In graph theory, graphs can be classified into directed and undirected graphs. A network represented by an undirected graph maintains the packet flow through its edges in the two directions; however, directed graphs have its edges with a directed flow from one end to the other and not the other way around. Any graph can be fully represented as an adjacency matrix. Whole information about a network is encoded in its adjacency matrix. Therefore, a graph can be constructed directly from its adjacency matrix. For unweighted undirected networks, the adjacency matrix **A** is a square symmetric matrix with zeros on the diagonal. The off-diagonal elements $A_{ij}$ are equal to 1 or 0 depending on whether the related nodes ($i$ and $j$) are connected or not[51]. For weighted networks, the elements can take other real values so that they represent the distance between the linked vertices. And for directed networks, the adjacency matrix is not symmetric anymore.

The Laplacian matrix is another representation of a graph that can be computed directly from the adjacency matrix to provide a vivid representation of the graph connectivity. For unweighted undirected graphs, the Laplacian matrix **L** is defined as[52];

$$\mathbf{L} = \mathbf{D} - \mathbf{A} \quad (1)$$

where **D** (the degree matrix) is a diagonal matrix in which the value of the element $d_{ii}$ equals the degree of the node $i$. The element $d_{ii}$ represents a crucial metric in graphs called local degree centrality. Consequently, the matrix **L** contains both of the edge information in its off-diagonal elements and the node degree in its diagonal.

Graph theory have been used in literature in analyzing connectivity of wireless networks. Connectivity probability of the network can be estimated, as a function of the node density and communication range, either indirectly by finding the probability of network isolation (partitioning) first[9], or directly by calculating the connectivity of the network graph through the Laplacian matrix. The Laplacian matrix differentiates between partitioned and connected graphs. A network graph is partitioned (clustered) if there exists a group of its vertices that is connected to form a sub-graph and is completely disconnected from other vertices in the same graph. The number of components[53] (clusters) in a graph/network can be determined using the eigenvalues of the Laplacian matrix as follows[54,55].

For a undirected graphs of $n$ vertices, the Laplacian matric **L** has $n$ real non-negative eigenvalues $\lambda_1 \leq \lambda_2 \leq \cdots \leq \lambda_n$, with $\lambda_1 = \cdots = \lambda_q = 0$ where q is the number of disjoint components in the graph. Hence, if the graph is connected, i.e., q = 1, then $0 = \lambda_1 < \lambda_2 \leq \cdots \leq \lambda_n$. In this case, $\lambda_2$ is called the algebraic connectivity of the graph or the spectral gap of the Laplacian matrix[54]. A detailed proof of this can be found in Merris[55] (see page 177 *ff.*).

Another advantage of representing VANETs as graphs is that other graph centrality metrics can also be used in network planning and analyses. These metrics include, but not restricted to closeness and betweenness centralities[56,57]. Closeness centrality measures how close a node is to other nodes within the network. This metric is vital for determining how long the data would take to travel from a specific node to other nodes in the network. Betweenness centrality, on the other hand, represents the number of shortest paths between different node pairs that pass through a given node. Therefore, the influence of this node on the data flow can be defined. These metrics were essential for designing carry-and forward routing strategies for ad-hoc networks, especially in delay-tolerant networking[12–15]. In contrast to these studies, we focus on the role of graph theory in VANET connectivity analysis from a physical-layer perspective, and then the adjacency matrix and algebraic connectivity are the key players in our proposed algorithms.

**III. Instantaneous Connectivity of Highway VANETs**

A highway VANET can be regarded as a one-dimensional network, in which, any two successive vehicles can exchange information if they fall within the communication range of one another. As we discuss in further detail below, the network is connected if all its node can communicate. Therefore, the necessary and sufficient condition for network connectivity is that the communication ranges are greater than the corresponding headway distances. The latter can be modelled as random variables.

The connectivity in a VANET can be determined as the probability that a (single-hop or multi-hop) communication path exists between any two vehicles in the considered road segment. Because of the continuous movement of nodes, which consistently affects the connectivity, we refer to the connectivity at a given moment as the instantaneous connectivity. Studying connectivity in wireless network is commonly performed either analytically or using numerical simulations. Numerical analyses are performed by conducting Monte-Carlo simulations. Monte-

Carlo simulations are carried out by repeating and averaging-out random topological realizations of the network under specific physical conditions (a fixed node degree, communication range, and certain spatial distribution of nodes). Each realization can be considered deterministic, and the average network behavior (mean connectivity probability in our case) is determined through finding the mean value of the connectivities of the different realizations[31,58,59].

To estimate the connectivity probability of a VANET using the unit disc model, we let the distance between any two consecutive vehicles $V_i$ and $V_{i+1}$ be denoted by the random variable $X_i$, $i = 1, 2, \ldots, N - 1$. In free-flow traffic, drivers are free to choose their own speeds, and cars can move independent of each other. Accordingly, the distance between any two cars is uncorrelated, and $X_i$'s are independent and identically distributed[17]. A VANET is connected if and only if $X_i$ is less than $R$, for all values of $i$. Knowing that the lane separation between vehicles is much less than the headway in free-flowing traffic, one can safely ignore the lane separation between vehicles in connectivity analysis[22,36]. The probability of connectivity $P_c$ is given[17] by (2).

$$P_c = \prod_{i=1}^{N-1} P(X_i < R) = \prod_{i=1}^{N-1} F_X(R) = F_X(R)^{N-1} \quad (2)$$

where $F_X(x)$ is the headway cumulative distribution function (CDF). In multi-lane free-flow traffic regime the headway distribution is commonly assumed to be exponential.

$$F_X(x) = (1 - e^{-\rho x})u(x) \quad (3)$$

where $u(x)$ is the Heaviside unit step function, and $\rho$ is the traffic density in vehicles per unit distance.

**IV. Graph-based Approaches for VANETs Connectivity Analysis**

We study the probability of the instantaneous connectivity of freeway VANETs. Without losing generalization, we limit our study to free-flowing traffic. This is mainly because in free-flow traffic phase ($\rho < 25$ vehicle/km)[60,61], the headway distances are large compared to those associated with congested traffic. In congested traffic regimes, the close distances between vehicles guarantee that they fall within the communication range of one another, in such regimes, the interference becomes a problem, but this falls outside the scope of this article.

Under the unit disc assumption, the communication channel is best described as symmetric. If a node is located within the coverage of its neighbor, this neighbor is certainly within the node's communication range as well. Hence, the network can be represented as an undirected unweighted graph with symmetric adjacency and Laplacian matrices. In the following, we propose two different graph-based approaches for estimating the VANET connectivity probability numerically while all vehicles employ the same fixed communication range.

## A. Laplacian Eigenvalue Approach

In this approach, we use the Laplacian matrix in numerically estimating the connectivity probability by averaging the connectivity over a graph ensemble. As discussed earlier, the Laplacian matrix can differentiate between partitioned and connected graphs. The number of components in a graph can be determined using the eigenvalues of the Laplacian matrix, which is calculated as

$$\mathbf{\Lambda} = \mathbf{U}^T \mathbf{L} \mathbf{U} \quad (4)$$

where $\mathbf{U}$ is the eigenvector matrix and $\mathbf{\Lambda}$ is a diagonal matrix that has the Laplacian eigenvalues in its diagonal. Over a large graph ensemble, we estimate the connectivity probability by the number of graphs whose second lowest eigenvalue is zero. For a given traffic density $\rho$ and a communication range R, the probability of connectivity can be estimated as in Algorithm 1.

---

**Algorithm 1** Laplacian-eigenvalue method for estimating the connectivity probability under fixed communication ranges.

---
1: Set the value of the communication range R
2: Set a spacing matrix $\mathbf{S}$ as an N × N zero matrix
3: Set *counter* = 0
4: **for** every iteration **do**
5:     Set a vector $\mathbf{y} \in \mathbb{R}^{N-1}$ with elements $y_i \sim \exp\left(\frac{1}{\rho}\right) \forall i$
6:     $S_{ij} = \sum_{m=i}^{j-1} y_m, \forall j > i, S_{ji} = S_{ij}$
7:     $\mathbf{A} = \mathbf{S} \leq R$
8:     $\mathbf{\Lambda} = \mathbf{U}^T(\mathbf{D} - \mathbf{A})\mathbf{U}$
9:     Set $\lambda_2$ as the second smallest value in the diagonal of $\mathbf{\Lambda}$
10:    **if** $\lambda_2 \neq 0$ **then**
11:        *counter* += 1
12: $P_c = \frac{counter}{\# \, iterations}$

---

To verify the correctness of this method, we consider a 10-km highway segment, in which the vehicles are free to select their speed independently on the other vehicles. The channel model is a unit disc with constant communication ranges R, which were set to three different values (500, 750, and 1000m) through simulations. To estimate the connectivity probability, Monte-Carlo simulations were performed, as discussed in Algorithm 1, and repeated for different values of the traffic density. The simulation results are then compared to the analytical connectivity of (3). Fig. 1 shows the match between the analytical and the Laplacian-eigenvalue-based simulation results.

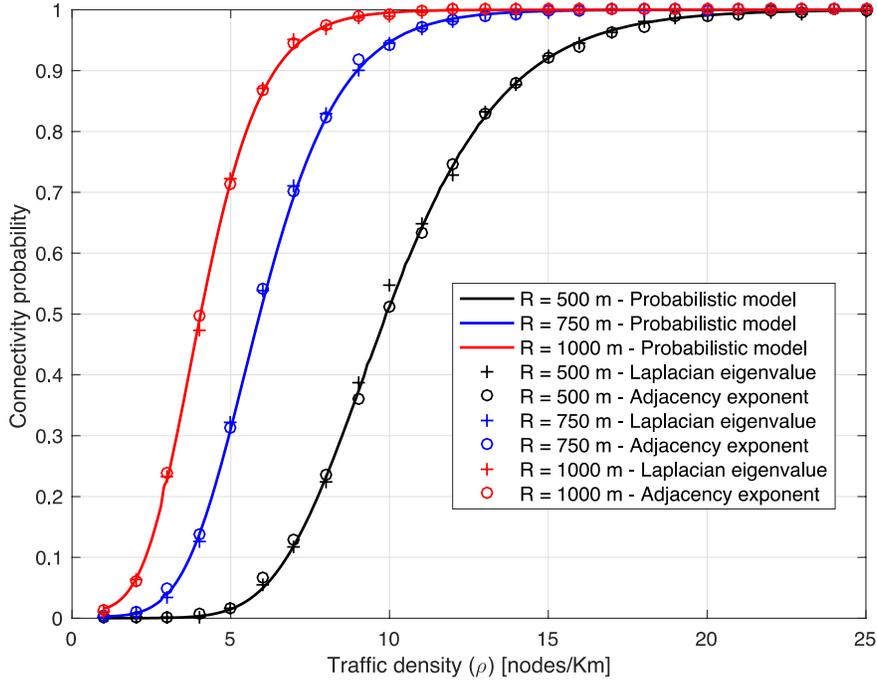

Fig. 1: Connectivity probability using the probabilistic (equation (3)) and the graph theory-based methods for different fixed communication ranges. The road segment length L = 10 km. Three different values of communication range R are used.

*B. Adjacency Exponent Approach*

A VANET can be thought of as (almost) a line graph as depicted in Fig 2. The main difference between line graphs and VANETs is that VANETs have additional edges that connect nodes to further non-consecutive neighbors. Then, the adjacency matrix can be of a great help in determining the network connectivity. While the elements of the adjacency matrix represent whether a connection exists between any two nodes, its square discloses the total number of walks of length two between any two nodes. In general, the $n^{th}$ exponent of the adjacency matrix counts the number of walks of length $n$ between any two nodes. In other words, the $ij^{th}$ element of the $n^{th}$ exponent of **A** equals the number of walks of length $n$ from $V_i$ to $V_j$. A line graph with $N$ vertices is connected if there exists a single walk of length $N-1$ between its first and last nodes. In a vehicular network, there would be more than one walk of length $N-1$ between the two end vehicles, if the network is connected, due to the extended links shown in Fig. 2b. Moreover, there would be other walks of a shorter length between the two intended nodes. Even though the shorter walks truthfully assert the connectivity, we stick to the walks of length $N-1$ to embrace the worst possible case of Fig. 2a, where the edges exist only between successor neighbors.

To estimate the connectivity probability using the adjacency matrix exponent method, we calculate the $(N-1)^{th}$ exponent of the matrix **A**, as described in Algorithm 2. The resultant matrix will be symmetric due to the communication channel symmetry. If the top right element or the bottom left one of the resultants is at least one, the network is declared connected. Simulations were held, and the results match the other methods, as depicted in Fig. 1.

---

**Algorithm 2** Adjacency-exponent method for estimating the connectivity probability under fixed communication ranges.

---
1: Set the value of the communication range R
2: Set a spacing matrix **S** as an N ×N zero matrix
3: Set *counter* = 0
4: **for** every iteration **do**
5:     Set a vector $\mathbf{y} \in \mathbb{R}^{N-1}$ with elements $y_i \sim \exp\left(\frac{1}{\rho}\right) \forall i$
6:     $S_{ij} = \sum_{m=i}^{j-1} y_m, \forall j > i, S_{ji} = S_{ij}$
7:     $\mathbf{A} = \mathbf{S} \leq R$
8:     **if** $(\mathbf{A}^{N-1})_{1,N} \neq 0$ **then**
9:        *counter* += 1
10: $P_c = \frac{counter}{\# iterations}$

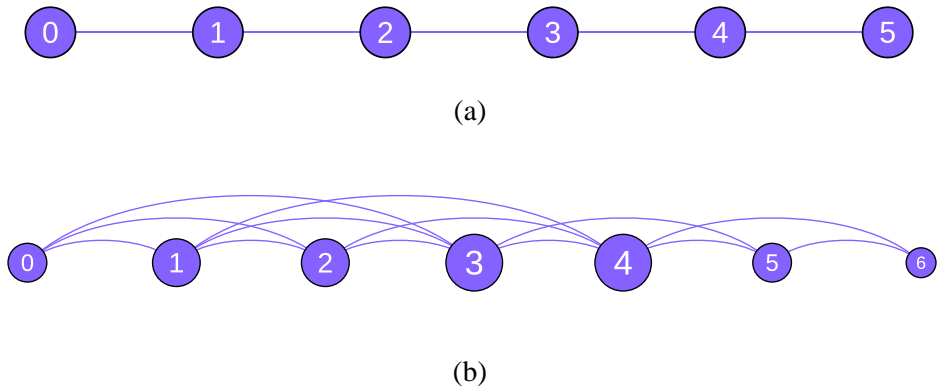

Fig. 2: Graph representation of (a) a line graph and (b) a VANET. The node size is proportional to the node degree.

## C. The Key Difference between the Two Approaches

Although the two methods estimate the VANET connectivity, a subtle difference can be inferred. The Laplacian eigenvalue method is dedicated to determining whether all vehicles are connected altogether as a single mesh. However, the adjacency exponent approach determines whether there exists a path between the vehicles located on each end of the road segment. In a unit disc model, where all vehicles utilize the same value of $R$, the two approaches match each other as provided in Fig. 1. As there would not be a path between the two end vehicles unless each vehicle is connected to its successor, providing connectivity to all vehicles. Fig. 3 depicts the case of a disconnected network under the fixed communication range assumption. Vehicle $V_4$ is too far away from $V_3$ to be located within its range; hence, it is farther away from the $V_3$'s predecessors, $V_1$ and $V_2$, and there is no way for $V_4$ to be connected to any of them. Therefore, the existence of a path between the two end vehicles relies on the fact that each vehicle should be connected to its close neighbors.

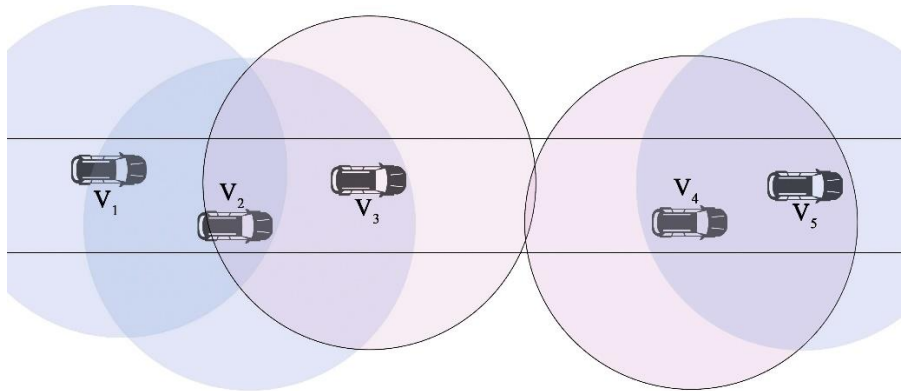

Fig. 3: Illustration of a VANET under a unit disc model.

In other channel models, where channel impairments are considered, $V_3$ and $V_4$ can be disconnected due to the communication channel between them, not the distance that separates them apart. However, at the same time, $V_4$ may be located on the communication range of $V_2$ or even $V_1$. In this case, the connection is guaranteed between the first and last vehicles, although there would be some isolated vehicles in between due to the randomness of the communication channels. Therefore, the two proposed approaches would deviate from each other representing two kinds of connectivity. Then, each approach is more convenient for typical applications. In end-to-end communications, where the network should be fully connected, the Laplacian eigenvalue approach can capture the network connectivity by ensuring that all vehicles are connected. The adjacency exponent approach, however, is suitable as a connectivity measure for routing, where it is crucial to forward the packet from one end of the road segment to the other regardless of the connectivity of each vehicle in the network. In this regard, the exponent can be even relaxed to a smaller value than $N-1$.

Additionally, when it comes to the computational complexity of both algorithms. The exponent method requires raising the adjacency matrix up to the $N-1$ power. Matrix multiplication has complexity of $O(n^3)$, and therefore, the algorithm has complexity of $O(n^4)$. On the other hand, determining the eigenvalues of the matrix, as required by the Laplacian-eigenvalues approach, has complexity $O(n^3)$.

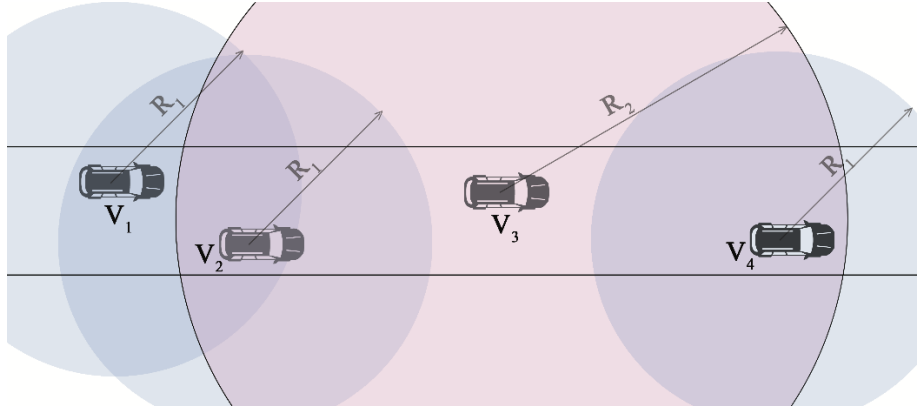

Fig. 4: Illustration of a VANET in a road segment that utilizes two communication ranges.

**V. Graph-based Connectivity Analysis for Random Communication Ranges**

It was recently suggested[62,63] that randomizing the transmit powers (varying the transmitting power of nodes according to a probability distribution, not dynamically, nor adaptively) in VANETs can improve some of their performance aspects. The randomness of the radio propagation environment caused by randomizing the transmit powers can keep the recurring interferences and collisions in safety dissemination applications as low as possible. Other communication benefits were presented by Kloiber el al[63]. Consequently, randomizing the transmit power directly results in randomizing the transmission range of VANETs nodes. Estimating the connectivity probability when different vehicles utilize different communication ranges is a more complicated problem. In this section we show how the method descried above can be used in such case.

*A. Two Different Fixed Communication Ranges*

We address the specific case when two values, $R_1$ and $R_2$, of the communication range are randomly assigned to different percentages of vehicles in the network. Fig. 4 illustrates a simple scenario of VANETs communications. The figure shows one major problem associated with utilizing two different communication ranges. As the figure shows, vehicle $V_2$ falls within the communication range of $V_3$, while the opposite is not true. This means that the network is no longer symmetric; $V_2$ can receive data packets being broadcasted from $V_3$, but not the other way around. The graph representing the network in such case has some directed edges.

Now that the variable communication range has resulted in a directed (or partly directed) network, the instantaneous connectivity of the network became asymmetric. Therefore, the estimation of probability of connectivity is understood to be for a single direction of communication, either upward (in the traffic flow direction), or downward (opposite to traffic flow). Therefore, one can define the probability of connectivity in this case as the probability of finding a communication path from the first to the last vehicles, or from the last to the first vehicles of the network. The problem of directed connectivity is, however, itself identical in the two directions; in every direction, the problem is to determine whether each vehicle is connected to its successor when the intervehicle spacings and random communication ranges follow the same distributions. Although the directed connectivity for a certain network is not guaranteed in both directions simultaneously, the probability of directed connectivity is the same regardless of the direction used in the estimation and whatever the upward and downward directions are. Thus, it is reasonable to narrow down the connectivity analysis to a certain direction, defined at the beginning of the analysis.

To extend the graph theory-based methods to enable connectivity analysis in such cases, a closer look at the adjacency and Laplacian matrices is required. Adjacency matrices of directed graphs are not symmetric. The upper and lower triangular parts of the adjacency matrix span the information of upward and downward directions, respectively. While only one direction is now required in connectivity analysis, one of the two triangular parts can be omitted. In our analysis, we will only consider the upward direction, and then all the elements under the diagonal of **A** are set to zero in Step 6 of Algorithms 1 and 2. The resultant adjacency matrix is still asymmetric, so is the Laplacian matrix.

With an asymmetric adjacency matrix, the adjacency exponent method, in Algorithm 2, would not experience any difficulty. For the new upper triangular matrix, **A**, the top right element of the $(N-1)^{th}$ exponent of **A** indicates the network connectivity. The other off-diagonal corner, however, will be always zero for omitting the backward connections. The connectivity probability, then, can be evaluated by counting the non-zero occurrence of the intended corner of the exponent matrices over a graph ensemble.

In contrast, the Laplacian eigenvalue method needs to be adapted to the considered case. The algebraic connectivity of directed (and mixed) graphs has various definitions and is harder to be determined directly from their adjacency and Laplacian matrices. Therefore, we propose enforcing the symmetry by completing the lower triangular part of the adjacency matrix. The new pseudo-adjacency matrix represents an artificial undirected network, where all the edges are bi-directional. If the algebraic connectivity of this graph is not zero, it implies that the connectivity exists in the two directions of that pseudo-network. That, in turn, admits the connectivity of the original network in the direction of our interest only.

Consequently, the composition of the adjacency matrix of Step 7 of Algorithm 1 should be altered as follows. After finding the spacing matrix **S**, for every vehicle $k$ with a communication range of $R_1$, the element $A_{kj}$ (for all $j > k$) is set to one if the corresponding spacing $S_{kj}$ is lower than $R_1$, otherwise it is set to zero. The same is done for the remaining vehicles with $R_2$ as their communication range. The elements under the diagonal of the adjacency matrix are left zeros for the adjacency exponent method, before the $(N-1)^{th}$ exponent matrix is evaluated. For the Laplacian eigenvalue method, the lower triangular part of the pseudo-adjacency matrix is instead filled to represent the pseudo backward connections. These backward elements would be set as $A_{jk} = A_{kj}$. The diagonal elements are always zeros.

For example, in the simple scenario illustrated in Fig. 4, we consider the network formed by the first three vehicles. The (asymmetric) adjacency matrix of the network consisting of the first three vehicles of the figure is given by.

$$A = \begin{bmatrix} 0 & 1 & 0 \\ 1 & 0 & 0 \\ 0 & 1 & 0 \end{bmatrix}$$

As proposed, one can re-express this in terms of two matrices, the upward and downward pseudo-adjacency matrices, which are, respectively, given by.

$$\tilde{A}_u = \begin{bmatrix} 0 & 1 & 0 \\ 1 & 0 & 0 \\ 0 & 0 & 0 \end{bmatrix}$$

$$\tilde{A}_d = \begin{bmatrix} 0 & 1 & 0 \\ 1 & 0 & 1 \\ 0 & 1 & 0 \end{bmatrix}$$

Accordingly, the upper and lower pseudo Laplacian matrices are given by.

$$\tilde{L}_u = \begin{bmatrix} 1 & -1 & 0 \\ -1 & 1 & 0 \\ 0 & 0 & 0 \end{bmatrix}$$

$$\tilde{L}_u = \begin{bmatrix} 1 & -1 & 0 \\ -1 & 2 & -1 \\ 0 & -1 & 1 \end{bmatrix}$$

These matrices imply that a downward communication path exists, while an upward path does not exist between $V_1$ and $V_3$.

Fig. 5 shows the resulting probability of connectivity of a VANET for different percentages of vehicles utilizing $R_2$ (Here, $R_1$ = 500 m, and $R_2$ = 1000 m). The figure shows that when approximately half of the vehicles utilized a communication range of 500 m, and the other half utilized a 1000 m range, the probability of connectivity curve was almost identical to that when all vehicles use a 750 m communication range.

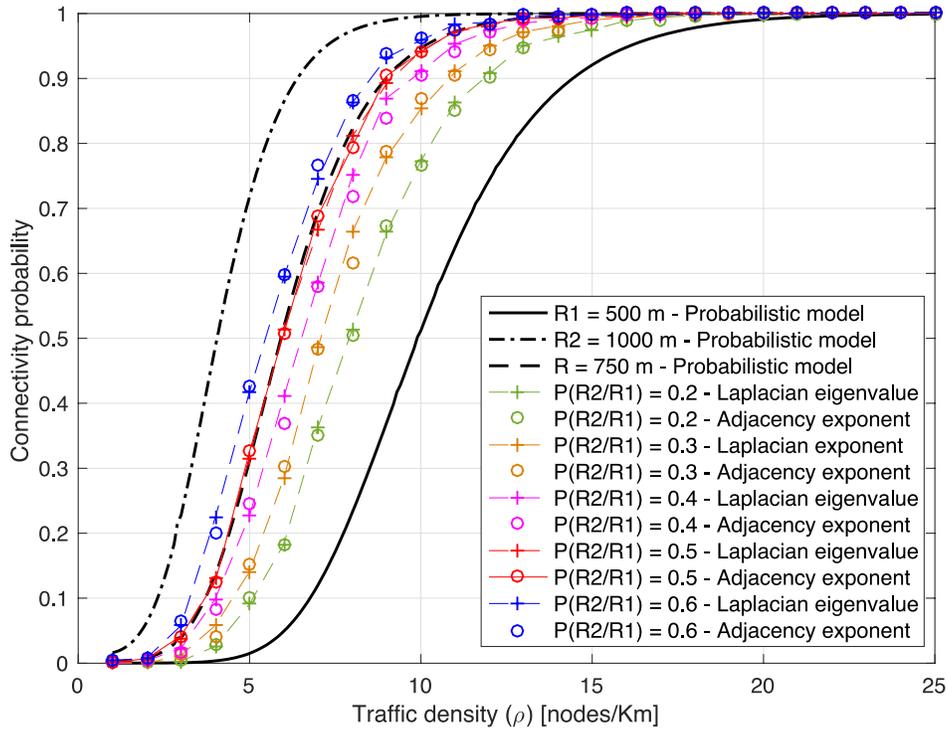

Fig. 5: Connectivity probability for a VANET with two communication ranges assigned randomly to different percentages of vehicles. L = 10 km, $R_1$ = 500 m, $R_2$ = 1000 m.

It is important to note that, for a given received SNR threshold, the transmit power is proportional to $R^\alpha$ (the communication range raised to the $\alpha$ power), where $\alpha$ is the path-loss exponent whose value depends on the propagation conditions. Therefore, one can deduce that under the same network conditions, to achieve similar VANET connectivity probability performance to that of networks with a fixed transmission range, a larger value of total used transmit power would be needed.

## B. Random Communication Ranges with Uniform Distribution

For the sake of generality, the procedures can be extended to cases where communication ranges follow a certain random distribution. For example, discrete random communication ranges with a uniform distribution can be assigned to the vehicles of the network, as suggested by Kloiber et al[62,63]. While such case would raise the complexity of acquiring a probabilistic model, it barely changes the graph-based methods. Fig. 6 depicts the case of assigning a discrete uniform-distributed communication range to each vehicle, with various mean values (i.e., 500, 750, and 1000 m) and a standard deviation of 100 m. The connectivity probability curves of randomized communication ranges match those when utilizing a fixed communication range that is equal to the mean of the employed communication range.

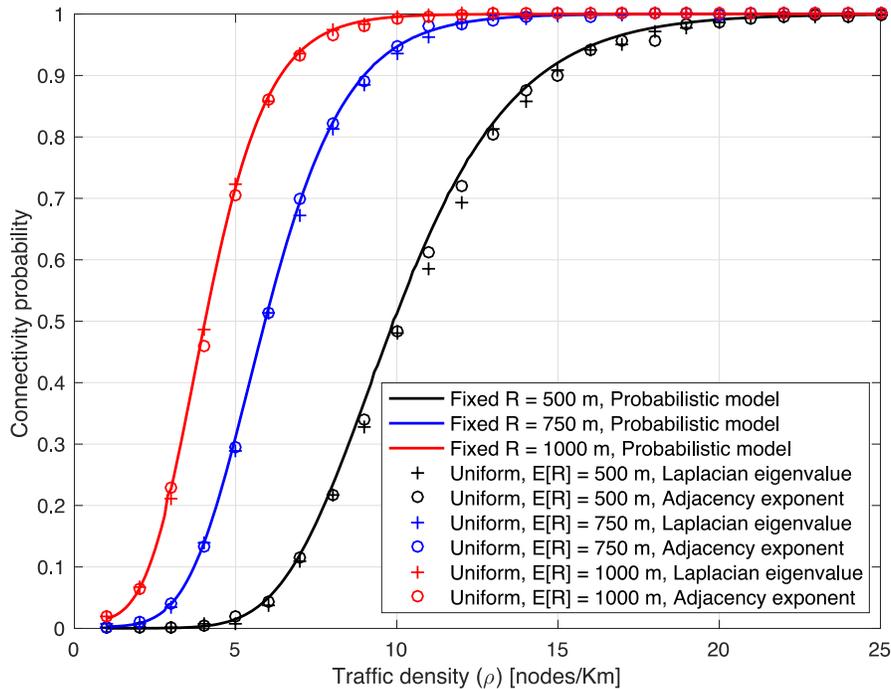

Fig. 6: Connectivity probability for a VANET with communication ranges assigned randomly based on a discrete uniform distribution with various mean values (L = 10 km).

Similarly, other more complicated scenarios, like fading and shadowing communication channels, can also be represented with graphs, and the same proposed procedures could precisely capture their connectivity probability. Moreover, different traffic models can also be handled without any difficulty. These scenarios may need sophisticated analysis before composing probabilistic models that capture their behavior, and it may be a challenging exercise[6]. This, in turn, indicates the significance of graph theory in the connectivity analysis and discloses its vital role in better understanding the connectivity of VANETs.

## VI. Conclusions

By representing a vehicular network as a graph, the latter's algebraic connectivity (i.e. the second smallest eigenvalue of the Laplacian matrix) and the exponent of its adjacency matrix hold information about the network connectivity. Based on these indicators, two different methods, with dedicated applications, were proposed and verified by comparing their simulation results to theoretically obtained ones. Next, graph-based methods proved their reliability by being extended to more sophisticated scenarios. The paper has focused on three scenarios; the scenario of a unit disc, two different communication ranges randomly employed by different vehicles, and discrete random communication ranges with uniform distribution. However, it can be extended to further scenarios. Besides its applications in designing routing paradigms for VANETs, graph theory shows superiority in connectivity analysis as well.

**Data availability statement**

The generated data that support the findings of this study are available from the corresponding author upon reasonable request.

**References**
1. Lu N, Cheng N, Zhang N, Shen X, Mark JW. Connected vehicles: Solutions and challenges. *IEEE Internet Things J*. 2014;1(4):289-299. doi:10.1109/JIOT.2014.2327587
2. Campolo C, Molinaro A, Vinel A, Zhang Y. Modeling and enhancing infotainment service access in vehicular networks with dual-radio devices. *Veh Commun*. 2016;6:7-16. doi:10.1016/j.vehcom.2016.10.001
3. Li Z, Song Y, Bi J. CADD: connectivity-aware data dissemination using node forwarding capability estimation in partially connected VANETs. *Wirel Networks*. 2019;25(1):379-398. doi:10.1007/s11276-017-1568-0
4. Busson A. Analysis and simulation of a message dissemination algorithm for VANET. *Int J Commun Syst*. 2011;24(9):1212-1229. doi:10.1002/dac.1226
5. Toor Y, Mühlethaler P, Laouiti A, De La Fortelle A. Vehicle ad hoc networks: Applications and related technical issues. *IEEE Commun Surv Tutorials*. 2008;10(3):74-88. doi:10.1109/COMST.2008.4625806
6. Naboulsi D, Fiore M. Characterizing the Instantaneous Connectivity of Large-Scale Urban Vehicular Networks. *IEEE Trans Mob Comput*. 2017;16(5):1272-1286. doi:10.1109/TMC.2016.2591527
7. Durrani S, Zhou X, Chandra A. Effect of vehicle mobility on connectivity of vehicular ad hoc networks. In: *IEEE Vehicular Technology Conference*. ; 2010. doi:10.1109/VETECF.2010.5594505
8. Khan Z, Fan P, Fang S. On the connectivity of vehicular Ad Hoc Network Under Various Mobility Scenarios. *IEEE Access*. 2017;5:22559-22565. doi:10.1109/ACCESS.2017.2761551
9. Dong J, Chen Q, Niu Z. Random graph theory based connectivity analysis in wireless sensor networks with rayleigh fading channels. In: *2007 Asia-Pacific Conference on Communications, APCC*. IEEE; 2007:123-126. doi:10.1109/APCC.2007.4433515

10. Madsen TK, Fitzek FHP, Prasad R. Impact of different mobility models on connectivity probability of a wireless ad hoc network. In: *2004 International Workshop on Wireless Ad-Hoc Networks*. ; 2005:120-124. doi:10.1109/iwwan.2004.1525554

11. Darwish TSJ, Abu Bakar K, Haseeb K. Reliable Intersection-Based Traffic Aware Routing Protocol for Urban Areas Vehicular Ad Hoc Networks. *IEEE Intell Transp Syst Mag*. 2018;10(1):60-73. doi:10.1109/MITS.2017.2776161

12. Yim J, Ahn H, Ko YB. The betweenness centrality based geographic routing protocol for unmanned ground systems. In: *ACM IMCOM 2016: Proceedings of the 10th International Conference on Ubiquitous Information Management and Communication*. New York, New York, USA: Association for Computing Machinery, Inc; 2016:1-4. doi:10.1145/2857546.2857622

13. Magaia N, Francisco AP, Pereira P, Correia M. Betweenness centrality in Delay Tolerant Networks: A survey. *Ad Hoc Networks*. 2015;33:284-305. doi:10.1016/j.adhoc.2015.05.002

14. Ahmed SH, Kang H, Kim D. Vehicular Delay Tolerant Network (VDTN): Routing perspectives. In: *2015 12th Annual IEEE Consumer Communications and Networking Conference, CCNC 2015*. Institute of Electrical and Electronics Engineers Inc.; 2015:898-903. doi:10.1109/CCNC.2015.7158095

15. Kang H, Ahmed SH, Kim D, Chung Y-S. Routing protocols for vehicular delay tolerant networks: a survey. *Int J Distrib Sens Networks*. 2015;11(3):325027.

16. Panichpapiboon S, Pattara-Atikom W. Connectivity requirements for self-organizing traffic information systems. *IEEE Trans Veh Technol*. 2008;57(6):3333-3340. doi:10.1109/TVT.2008.929067

17. Cheng L, Panichpapiboon S. Effects of intervehicle spacing distributions on connectivity of VANET: A case study from measured highway traffic. *IEEE Commun Mag*. 2012;50(10):90-97. doi:10.1109/MCOM.2012.6316781

18. Yousefi S, Altman E, El-Azouzi R, Fathy M. Analytical model for connectivity in vehicular Ad Hoc networks. *IEEE Trans Veh Technol*. 2008;57(6):3341-3356. doi:10.1109/TVT.2008.2002957

19. C. Neelakantan P, V. Babu A. Network Connectivity Probability of Linear Vehicular Ad Hoc Networks on Two-Way Street. *Commun Netw*. 2012;04(04):332-341. doi:10.4236/cn.2012.44038

20. Baltzis KB. On the effect of channel impairments on VANETs performance. *Radioengineering*. 2010;19(4):689-694.

21. Chen C, Du X, Pei Q, Jin Y. Connectivity analysis for free-flow traffic in vanets: A statistical approach. *Int J Distrib Sens Networks*. 2013;2013(1):598946. doi:10.1155/2013/598946

22. Nagel R. The effect of vehicular distance distributions and mobility on VANET communications. In: *IEEE Intelligent Vehicles Symposium, Proceedings*. ; 2010:1190-1194. doi:10.1109/IVS.2010.5547971

23. Abuelenin SM, Abul-Magd AY. Effect of minimum headway distance on connectivity of VANETs. *AEU - Int J Electron Commun*. 2015;69(5):867-871. doi:10.1016/j.aeue.2015.01.011

24. Abuelenin SM, Abul-Magd AY. Studying Connectivity Probability and Connection Duration in Freeway VANETs. In: *Studies in Systems, Decision and Control*. Vol 242. Springer International Publishing; 2020:27-38. doi:10.1007/978-3-030-22773-9_3

25. Hoque MA, Hong X, Dixon B. Efficient multi-hop connectivity analysis in urban vehicular networks. *Veh Commun*. 2014;1(2):78-90. doi:10.1016/j.vehcom.2014.04.002

26. Elaraby S, Abuelenin SM. Fading Improves Connectivity in Vehicular Ad-hoc Networks. *arXiv Prepr arXiv191005317*. October 2019.

27. Zhang L, Cai L, Pan J, Tong F. A new approach to the directed connectivity in two-dimensional lattice networks. *IEEE Trans Mob Comput*. 2014;13(11):2458-2472. doi:10.1109/TMC.2014.2314128

28. Xiao H, Zhang Q, Ouyang S, Chronopoulos AT. Connectivity Probability Analysis for VANET Freeway Traffic Using a Cell Transmission Model. *IEEE Syst J*. June 2020:1-10. doi:10.1109/jsyst.2020.3001938

29. Chen G, Wang X, Li X. *Fundamentals of Complex Networks: Models, Structures and Dynamics*.; 2015. doi:10.1002/9781118718124

30. Haenggi M. *Stochastic Geometry for Wireless Networks*. Vol 9781107014695. Cambridge University Press; 2009. doi:10.1017/CBO9781139043816

31. Dung LT, Choi SG. Connectivity analysis of cognitive radio Ad-Hoc networks with multi-pair primary networks. *Sensors (Switzerland)*. 2019;19(3). doi:10.3390/s19030565

32. Takabe S, Wadayama T. Approximation Theory for Connectivity of Ad Hoc Wireless Networks with Node Faults. *IEEE Wirel Commun Lett*. 2019;8(4):1240-1243. doi:10.1109/LWC.2019.2912610

33. Babu A V., Muhammed Ajeer VK. Analytical model for connectivity of vehicular ad hoc networks in the presence of channel randomness. *Int J Commun Syst*. 2013;26(7):927-946. doi:10.1002/dac.1379

34. Andrews JG, Ganti RK, Haenggi M, Jindal N, Weber S. A primer on spatial modeling and analysis in wireless networks. *IEEE Commun Mag*. 2010;48(11):156-163. doi:10.1109/MCOM.2010.5621983

35. Abuelenin SM, Abul-Magd AY. Corrigendum to "Effect of minimum headway distance on connectivity of VANETs" [AEU – Int. J. Electron. Commun. 69(5) (2015) 867–871]. *AEU - Int J Electron Commun*. 2017;83:566. doi:10.1016/j.aeue.2017.10.036

36. Abuelenin SM, Abul-Magd AY. Empirical study of traffic velocity distribution and its effect on vanets connectivity. In: *2014 International Conference on Connected Vehicles and Expo (ICCVE)*. IEEE; 2014:391-395. doi:10.1109/ICCVE.2014.7297577

37. Yan G, Olariu S. A probabilistic analysis of link duration in vehicular ad hoc networks.

*IEEE Trans Intell Transp Syst*. 2011;12(4):1227-1236. doi:10.1109/TITS.2011.2156406

38. Kwon S, Kim Y, Shroff NB. Analysis of Connectivity and Capacity in 1-D Vehicle-to-Vehicle Networks. *IEEE Trans Wirel Commun*. 2016;15(12):8182-8194. doi:10.1109/TWC.2016.2613078

39. Qu X, Liu E, Wang R, Ma H. Complex Network Analysis of VANET Topology with Realistic Vehicular Traces. *IEEE Trans Veh Technol*. 2020;69(4):4426-4438. doi:10.1109/TVT.2020.2976937

40. Rodríguez JM, Sigarreta JM. Spectral properties of geometric-arithmetic index. *Appl Math Comput*. 2016;277:142-153. doi:10.1016/j.amc.2015.12.046

41. Swain RR, Dash T, Khilar PM. An effective graph-theoretic approach towards simultaneous detection of fault(s) and cut(s) in wireless sensor networks. *Int J Commun Syst*. 2017;30(13):e3273. doi:10.1002/dac.3273

42. Devi RK, Murugaboopathi G. An efficient clustering and load balancing of distributed cloud data centers using graph theory. *Int J Commun Syst*. 2019;32(5):e3896. doi:10.1002/dac.3896

43. Tong C, Niu JW, Qu GZ, Long X, Gao XP. Complex networks properties analysis for mobile ad hoc networks. *IET Commun*. 2012;6(4):370-380. doi:10.1049/iet-com.2010.0651

44. Gao B, Yang Y, Ma H. A new distributed approximation algorithm for constructing minimum connected dominating set in wireless ad hoc networks. *Int J Commun Syst*. 2005;18(8):743-762. doi:10.1002/dac.726

45. Meng Y, Dong Y, Liu X, Zhao Y. An Interference-Aware Resource Allocation Scheme for Connectivity Improvement in Vehicular Networks. *IEEE Access*. 2018;6:51319-51328. doi:10.1109/ACCESS.2018.2867745

46. Kawahigashi H, Terashima Y, Miyauchi N, Nakakawaji T. Modeling ad hoc sensor networks using random graph theory. In: *2005 2nd IEEE Consumer Communications and Networking Conference, CCNC2005*. Vol 2005. ; 2005:104-109. doi:10.1109/ccnc.2005.1405152

47. Yano A, Wadayama T. Probabilistic analysis of the network reliability problem on random graph ensembles. In: *IEICE Transactions on Fundamentals of Electronics, Communications and Computer Sciences*. Vol E99A. Maruzen Co., Ltd.; 2016:2218-2225. doi:10.1587/transfun.E99.A.2218

48. Zhao J. *Minimum Node Degree and K-Connectivity in Wireless Networks with Unreliable Links*.

49. Li J, Andrew LLH, Foh CH, Zukerman M, Chen HH. Connectivity, coverage and placement in wireless sensor networks. *Sensors*. 2009;9(10):7664-7693. doi:10.3390/s91007664

50. Naboulsi D, Fiore M. On the instantaneous topology of a large-scale urban vehicular network: The cologne case. In: *Proceedings of the International Symposium on Mobile Ad*

*Hoc Networking and Computing (MobiHoc)*. New York, New York, USA: ACM Press; 2013:167-175. doi:10.1145/2491288.2491312

51. Abuelenin SM, Abul-Magd AY. Effect of unfolding on the spectral statistics of adjacency matrices of complex networks. *Procedia Comput Sci*. 2012;12:69-74. doi:10.1016/j.procs.2012.09.031

52. Weisstein EW. "Algebraic Connectivity." From MathWorld--A Wolfram Web Resource. https://mathworld.wolfram.com/AlgebraicConnectivity.html. Accessed March 24, 2020.

53. Merris R. Laplacian matrices of graphs: a survey. *Linear Algebra Appl*. 1994;197-198(C):143-176. doi:10.1016/0024-3795(94)90486-3

54. Chen G, Wang X, Li X. *Fundamentals of Complex Networks*. Singapore: John Wiley & Sons Singapore Pte. Ltd; 2014. doi:10.1002/9781118718124

55. Merris R. *Graph Theory*. Hoboken, NJ, USA: John Wiley & Sons, Inc.; 2000. doi:10.1002/9781118033043

56. Pallis G, Katsaros D, Dikaiakos MD, Loulloudes N, Tassiulas L. On the structure and evolution of vehicular networks. In: *Proceedings - IEEE Computer Society's Annual International Symposium on Modeling, Analysis, and Simulation of Computer and Telecommunications Systems, MASCOTS*. ; 2009:502-511. doi:10.1109/MASCOT.2009.5366230

57. Akabane AT, Immich R, Pazzi RW, Madeira ERM, Villas LA. Distributed egocentric betweenness measure as a vehicle selection mechanism in VANETs: A performance evaluation study. *Sensors (Switzerland)*. 2018;18(8). doi:10.3390/s18082731

58. Lu W, Bao Y, Sun X, Wang Z. Performance evaluation of inter-vehicle communication in a unidirectional dynamic traffic flow with shockwave. In: *2009 International Conference on Ultra Modern Telecommunications and Workshops*. ; 2009. doi:10.1109/ICUMT.2009.5345416

59. Zhe W, Wei L, Wenlong J. A study on information throughput of inter-vehicle communications in a unidirectional traffic stream. In: *Proceedings - 2009 WRI International Conference on Communications and Mobile Computing, CMC 2009*. Vol 2. ; 2009:396-401. doi:10.1109/CMC.2009.226

60. Ayres TJ, Li L, Schleuning D, Young D. Preferred time-headway of highway drivers. In: *IEEE Conference on Intelligent Transportation Systems, Proceedings, ITSC*. ; 2001:826-829. doi:10.1109/itsc.2001.948767

61. Abuelenin SM, Abul-Magd AY. Moment Analysis of Highway-Traffic Clearance Distribution. *Intell Transp Syst IEEE Trans*. 2015;16(99):1-8. doi:10.1109/TITS.2015.2412117

62. Kloiber B, Harri J, Strang T. Dice the TX power-improving awareness quality in VANETs by random transmit power selection. In: *IEEE Vehicular Networking Conference, VNC*. ; 2012:56-63. doi:10.1109/VNC.2012.6407445

63. Kloiber B, Härri J, Strang T, Sand S, García CR. Random Transmit Power Control for